# Photon statistics of amplified spontaneous emission

I. V. Doronin,[1,2] E. S. Andrianov,[1,2] A. A. Zyablovsky,[1,2] A. A. Pukhov,[1-3] Yu. E. Lozovik,[1,2,4] A. P. Vinogradov,[1-3] A. A. Lisyansky,[5,6]

[1]Dukhov Research Institute of Automatics, 127055, Moscow, Russia
[2]Moscow Institute of Physics and Technology, 141700, Moscow region, Russia
[3]Institute for Theoretical and Applied Electromagnetics, 125412, Moscow, Russia
[4]Institute of Spectroscopy RAS, 142190, Troitsk, Moscow, Russia
[5]Department of Physics, Queens College of the City University of New York, Queens, New York, 11367, USA
[6]The Graduate Center of the City University of New York, New York, 10016, USA

Developments in quantum technologies lead to new applications that require radiation sources with specific photon statistics. A widely used Poissonian statistics are easily produced by lasers; however, some applications require super- or sub-Poissonian statistics. Statistical properties of a light source are characterized by the second-order coherence function $g^{(2)}(0)$. This function distinguishes stimulated radiation of lasers with $g^{(2)}(0) \approx 1$ from light of other sources. For example, $g^{(2)}(0) = 2$ for black-body radiation, and $g^{(2)}(0) = 0$ for single-photon emission. One of the applications requiring super-Poissonian statistics ($g^{(2)}(0) > 1$) is ghost imaging with thermal light. Ghost imaging also requires light with a narrow linewidth and high intensity. Currently, rather expensive and inefficient light sources are used for this purpose. In the last year, a superluminescent diode based on amplified spontaneous emission (ASE) has been considered as a new light source for ghost imaging. Even though ASE has been widely studied, its photon statistics has not been settled - there are neither reliable theoretical estimates of the second-order coherence function nor unambiguous experimental data. Our computer simulation clearly establishes that coherence properties of light produced by ASE are similar to that of a thermal source with $g^{(2)}(0) \approx 2$ independent of pump power. This result manifests the fundamental difference between ASE and laser radiation.

## 1. INTRODUCTION

The development of quantum technologies has led to the emergence of a number of applications that require radiation sources with specific photon statistic [1-4].

The statistical properties of radiation are characterized by the second-order coherence function $g^{(2)}(0) = \langle \hat{I}(t)\hat{I}(t+\tau) \rangle / \langle \hat{I}(t) \rangle^2$, where $\hat{I}(t)$ is the operator of radiation intensity. Coherent laser light has Poissonian photon distributions with $g^{(2)}(0) = 1$, while black-body and single-photon sources have super- and sub-Poissonian photon distributions for which $g^{(2)}(0)$ is equal to 2 and 0, respectively.

An example of the application that needs the specific statistics of photons is ghost imaging with pairs of entangled photons [5, 6], for which light sources with sub-Poissonian sources are used. One of the photons forms an image, while the second one is used for the control to separate the signal from noise [5, 7-9]. Obtaining light with such statistics requires sources based on parametric down-conversion [5, 7-11] which are expensive and difficult to manufacture. Recently it has been shown that the desirable light source could be replaced with a source that emits light with the super-Poissonian distribution in which photon pairs are created randomly [6]. The rest of radiation is spurious and is treated as noise. The coherence length of radiation for ghost imaging must be greater than the size of the object studied [10]. Therefore, the source should have a narrow linewidth and a high intensity [10, 12, 13].

Below, we focus our attention on light sources having a super-Poissonian statistics of the photons for which $g^{(2)}(0) > 1$. Currently, either lasers with rotating grounded glasses [14-16] with the value of $g^{(2)}(0)$ between 1.25 and 1.9 or incoherent lamps with a frequency filter [17, 18] that have the value of $g^{(2)}(0)$ of about 1.05 are used as such radiation sources. However, these sources have small efficiencies and broad radiation patterns [19].

In last the year, the superluminescent diode based on amplified spontaneous emission (ASE), which has narrow linewidth and high intensity, is considered as a new light source for ghost imaging [12]. Employing ASE has also been stimulated by applications of new luminescent materials with high gain and quantum yield such as organo-lead halide [20, 21] and perovskites [22, 23]. The low-cost production of single nanowire ASE sources based on these materials and the broad range of light emission wavelengths make such sources attractive for optoelectronic and optical storage devices [24].

ASE or super-luminescence can be produced by an active medium pumped by incoherent radiation. When a luminescent material is pumped, one of the excited atoms spontaneously emits a photon. Before leaving the active medium, this photon triggers stimulated emission of inverted atoms on its path [23, 25, 26]. Thus, ASE is an interplay of lasing and black-body radiation.

ASE sources have many common features with conventional lasers. Due to system losses, the amplification of the electromagnetic (EM) waves propagating through an active medium takes place only when the pump rate exceeds a certain value called the compensation threshold. Above the threshold, the amplification due to induced transitions in the inverted medium overcomes the loss. Further increase in the pump rate results in the exponential increase of the intensity $I(t)$ of EM waves propagating through the inverted medium. At high intensities of the EM field, due to the saturation of the population inversion of the active medium, the gain starts decreasing and the growth of $I(t)$ becomes linear. Thus, similar to lasers, the input-output curve of an ASE source is s-shaped and there are three regimes of the pump rates: (1) below the compensation threshold, in which the spontaneous emission of atoms plays the key role and EM waves decay as the wave propagates through the active medium; (2) above the saturation threshold, when the growth of the intensity of the EM field with the pump power is linear; and (3) the intermediate regime, in which an increase in the pump rate results in the exponential increase of $I(t)$.

At the transition frequency of inverted atoms, the gain of the active medium has a maximum. Thus, the amplification of the EM at neighboring frequencies is smaller. This results in narrowing of a spectral linewidth of ASE sources above the compensation threshold.

ASE light sources also exhibit properties that differ from the properties of conventional lasers. Unlike lasing, which demonstrates the frequency pulling, the frequency of ASE coincides with the transition frequency of an active medium. This happens because in ASE, there is no feedback and it is not self-oscillation.

Statistical properties of ASE, however, are not well-studied because for a wide spectral range of radiation the determination of the distribution of emitted photons is a complicated problem [19]. Recent experiments have yield contradicting results. In Ref. [27] it was shown that above the threshold, $g^{(2)}(0)$ is about unity, which characterizes ASE as lasing, whereas Refs. [12, 28] reported that $g^{(2)}(0) = 2$ characteristic to black-body radiation.

In this paper, we investigate the dependence of the statistical properties of ASE on the pump rate. Using computer simulation, we demonstrate that ASE sources have the super-Poissonian distribution of photons with $g^{(2)}(0) \approx 2$ for any value of the pump rate.

## 2. THE MODEL FOR THE LIGHT INTERACTION WITH ACTIVE MEDIUM

To describe the interaction between an EM field and atoms of a gain medium we use Heisenberg-Langevin equation employing the Jaynes-Cummings Hamiltonian in the rotating wave approximation:

$$\hat{H} = \sum_i \hbar \omega_i \hat{a}_i^\dagger \hat{a}_i + \sum_j \hbar \omega_{TLS} \hat{\sigma}_j^\dagger \hat{\sigma}_j + \sum_{i,j} \hbar \Omega_{ij} \left( \hat{a}_i^\dagger \hat{\sigma}_j + \hat{a}_i \hat{\sigma}_j^\dagger \right) + \hat{H}_R + \hat{H}_{SR}, \tag{1}$$

where $\hat{a}_i^\dagger$ and $\hat{a}_i$ are creation and annihilation operators for the $i$-th mode of the EM field. These operators satisfy the commutation relation $\left[ \hat{a}_{i1}, \hat{a}_{i2}^\dagger \right] = \delta_{i1,i2}$. The electric field operator at the point $\mathbf{r}_j$ is expressed as $\hat{\mathbf{E}}(\mathbf{r}_j) = \sum_i \mathbf{E}_i(\mathbf{r}_j) \hat{a}_i$, where $\mathbf{E}_i(\mathbf{r}) = \sqrt{4\pi \hbar \omega_i / V} \cos(k_i \mathbf{r})$ is the electric field "per one photon" of $i$-th mode and $V$ is the volume of the system. Atoms of the active medium are described as two-level systems (TLSs) with ground, $|g\rangle$ and excited, $|e\rangle$, states, $\hat{\sigma}_j^+ = |e_j\rangle\langle g_j|$, $\hat{\sigma}_j = |g_j\rangle\langle e_j|$, and $\hat{D}_j = |e_j\rangle\langle e_j| - |g_j\rangle\langle g_j|$ are raising, lowering and population inversion operators of the $j$-th TLS, respectively, and $\omega_{TLS}$ is its transition frequency. The third term in Eq. (1) describes the interaction between field modes and dipole moments of TLSs. The coupling constant $\Omega_{ij}$ (the Rabi frequency) is equal to $-\mathbf{E}_i(\mathbf{r}_j)\mathbf{d}_j / \hbar$, where $\mathbf{r}_j$ is the position of the $j$-th TLS and $\mathbf{d}_j = \langle e|e\mathbf{r}|g \rangle_j$ is the matrix element of its dipole moment. $\hat{H}_R$ is a sum of the Hamiltonians of the reservoirs $\hat{H}_R = \hat{H}_{Ra} + \sum_j \hat{H}_{R\sigma}^{(j)} + \sum_j \hat{H}_{RPump}^{(j)}$, where $\hat{H}_{Ra} = \sum_j \hbar \omega_j \hat{b}_j^\dagger \hat{b}_j$ describes the reservoir of phonons in waveguide walls, $\hat{H}_{R\sigma}^{(j)} = \sum_n \hbar \omega_{jn} \hat{c}_{jn}^\dagger \hat{c}_{jn}$ is the Hamiltonian of the reservoir of phonons in the active medium coupled with the $j$-th atom, the operators $\hat{b}_j$, $\hat{c}_{jn}$ and $\hat{b}_j^\dagger$, $\hat{c}_{jn}^\dagger$ are annihilation and creation operators of phonons. We assume that every atom of the active medium couples with its own phonon reservoir [29]. This approximation is valid when the coherence length of the phonons is less than the distance between atoms of the active medium.

To describe pumping of the active medium atoms, we introduce reservoirs of auxiliary TLSs having negative temperature. The *j*-th reservoir interacts with the *j*-th atom, and is described by the Hamiltonian $\hat{H}_{RPump}^{(j)} = \sum_n \hbar \omega_s \hat{s}_{jn}^\dagger \hat{s}_{jn}$, where $\hat{s}_{jn}$ and $\hat{s}_{jn}^\dagger$ are lowering and raising population inversion operators for TLSs in the reservoir with negative temperature.

The Hamiltonian that describes the interaction of the system with these reservoirs is the sum $H_{SR} = \hat{H}_{SRa} + \sum_j \hat{H}_{SRdeph}^{(j)} + \sum_j \hat{H}_{SRD}^{(j)} + \sum_j \hat{H}_{SRPump}^{(j)}$. The Hamiltonian $\hat{H}_{SRa} = \sum_{ij} v_{ij} \left( \hat{a}_i \hat{b}_j^\dagger + \hat{a}_i^\dagger \hat{b}_j \right)$ describes the interaction of phonons in the waveguide walls with photons. The Hamiltonians $\hat{H}_{SRdeph}^{(j)} = \sum_n \kappa_{jn} \left( \hat{c}_{jn} + \hat{c}_{jn}^\dagger \right) \hat{\sigma}_j^\dagger \hat{\sigma}_j$ and $\hat{H}_{SRD}^{(j)} = \sum_n \beta_{jn} \left( \hat{c}_{jn} \hat{\sigma}_j^\dagger + \hat{c}_{jn}^\dagger \hat{\sigma}_j \right)$ describe the elastic and inelastic interactions between *j*-th atom and phonons in the active medium, $\hat{H}_{SRPump}^{(j)} = \sum_j \eta_{jn} \left( \hat{\sigma}_j \hat{s}_{jn}^\dagger + \hat{\sigma}_j^\dagger \hat{s}_{jn} \right)$ is the Hamiltonian of the interaction between *j*-th atom of the active medium and TLSs in the reservoir with negative temperature.

To obtain the Heisenberg equations of motion for the system operators, we eliminate the reservoir variables by using Born and Markov approximations [30]. In the Born approximation, it is assumed that the reservoir is much greater than the system, and therefore, one can neglect an influence of the system on the reservoir [30]. The Markov approximation assumes that time-scales, characterizing fluctuations in the reservoir, are much shorter than the relaxation time of the system [30]. This assumption allows one to derive the fluctuation-dissipation theorem [30]. For a reservoir in the thermal equilibrium at room temperature, the characteristic time may be estimated as $\tau_R \sim \hbar/kT \sim 10^{-14} s$ [30] for a typical gain medium (e.g., for dye R101), the relaxation time is about $10^{-13} s$ that is greater than $\tau_R$, thus, the Markov approximation is applicable.

After the elimination of the reservoir variables, the relaxation and fluctuation (noise) terms appear in the equations. Note that the corresponding terms are connected via Fluctuation-Dissipative theorem. In particular, for the operators $\hat{a}_i^\dagger$, $\hat{a}_i$, $\hat{\sigma}_j^\dagger$, $\hat{\sigma}_j$, and $\hat{D}_j$ we arrive at the following system of equations (see also [29, 30]):

$$\frac{d}{dt}\hat{a}_i = \left(-\gamma_a/2 - i\Delta_i\right)\hat{a}_i - i\sum_j \Omega_{ij}\hat{\sigma}_j + \hat{F}_{ai}, \quad (2)$$

$$\frac{d}{dt}\hat{a}_i^\dagger = \left(-\gamma_a/2 + i\Delta_i\right)\hat{a}_i^\dagger + i\sum_j \Omega_{ij}\hat{\sigma}_j^\dagger + \hat{F}_{a^\dagger i}, \quad (3)$$

$$\frac{d}{dt}\hat{\sigma}_j = -\hat{\sigma}_j(\gamma_p + \gamma_D + \gamma_{deph})/2 + i\sum_i \Omega_{ij}\hat{a}_i\hat{D}_j + \hat{F}_{\sigma j}, \quad (4)$$

$$\frac{d}{dt}\hat{\sigma}_j^\dagger = -\hat{\sigma}_j^\dagger(\gamma_p + \gamma_D + \gamma_{deph})/2 - i\sum_i \Omega_{ij}\hat{a}_i^\dagger\hat{D}_j + \hat{F}_{\sigma^\dagger j}, \quad (5)$$

$$\frac{d}{dt}\hat{D}_j = -(\gamma_p + \gamma_D)\left(\hat{D}_j - D_0\right) + 2i\sum_i \Omega_{ij}\left(\hat{a}_i^\dagger\hat{\sigma}_j - \hat{a}_i\hat{\sigma}_j^\dagger\right) + \hat{F}_{Dj}, \quad (6)$$

where $\gamma_a$, $\gamma_D$, and $\gamma_{deph}$ are the relaxation rates of field modes $\hat{a}_i$, the population inversion $D$, and dipole moments of active atoms $\sigma$; $\gamma_p$ is the atom pump rate; $D_0 = (\gamma_p - \gamma_D)/(\gamma_p + \gamma_D)$ is unsaturated population inversion, and $\hat{F}(t)$ are the Langevin noise operators [29, 30]. Since each atom of the active medium interacts with its own environment described by the corresponding phonon reservoir, correlation functions of noise operators of different atoms are assumed to be equal to zero [29], i.e.

$$\left\langle \hat{F}_i(t) \hat{F}_j(t') \right\rangle \sim \delta_{ij}. \tag{7}$$

The correlation functions of the noise operators of Eqs. (2)-(6) are

$$\left\langle \hat{F}_{\sigma j}(t) \hat{F}_{\sigma j}(t') \right\rangle = i \left\langle \hat{\sigma}_j^{st} \sum_k \hat{a}_k^{st} \Omega_{kj} \right\rangle \delta(t - t'), \tag{8}$$

$$\left\langle \hat{F}_{\sigma^\dagger j}(t) \hat{F}_{\sigma^\dagger j}(t') \right\rangle = -i \left\langle \sum_k \left(\hat{\sigma}_j^\dagger\right)^{st} \left(\hat{a}_k^\dagger\right)^{st} \Omega_{kj} \right\rangle \delta(t - t'), \tag{9}$$

$$\left\langle \hat{F}_{\sigma^\dagger j}(t) \hat{F}_{\sigma j}(t') \right\rangle = \frac{1}{2}\left[\gamma_p + \gamma_{ph}\left(1 + \left\langle \hat{D}_j^{st} \right\rangle\right)\right]\delta(t - t'), \tag{10}$$

$$\left\langle \hat{F}_{Dj}(t) \hat{F}_{Dj}(t') \right\rangle = \left[(\gamma_D + \gamma_p)\left(D_0 - \left\langle \hat{D}_j^{st} \right\rangle\right) + 2\sum_k \Omega_{kj}\left(\left\langle \left(\hat{a}_k^\dagger\right)^{st} \hat{\sigma}_j^{st} - \hat{a}_k^{st}\left(\hat{\sigma}_j^\dagger\right)^{st} \right\rangle\right)\right]\delta(t - t'), \tag{11}$$

where $\hat{D}_j^{st}$, $\hat{\sigma}_j^{st}$, and $\hat{a}_j^{st}$ are stationary solutions of Eqs. (2)–(6) without noise terms. The average values of noises included in the equations for the annihilation and creation operators of the EM field are negligibly small in comparison with the average values of the noises in the equations for the atoms. Therefore, below we do not take them into account $\hat{F}_{ai}$ and $\hat{F}_{a^\dagger i}$.

Due to the presence of non-commuting noise operators, Eqs. (2)–(6) are difficult to solve. In the most of ASE experiments, the number $N_c$ of inverted atoms in a subwavelength volume of the size $L \sim \lambda/10$ ($\lambda$ is the transition wavelength of gain medium atoms) is large. For example, in a volume of the size about 100 nm, $N_c$ is of the order of $10^3$. This allows us to use the method of large cells. We divide the whole volume into cells of the size $0.1\lambda$ and transition to the mean operators: $\hat{J}_k^\dagger = \sum_{j \in cell_k} \hat{\sigma}_j^\dagger / N_c$, $\hat{J}_k = \sum_{j \in cell_k} \hat{\sigma}_j / N_c$, and $\hat{J}_{D,k} = \sum_{j \in cell_k} \hat{D}_j / N_c$, where $N_c$ is the number of atoms in a cell. Following the method of the system size expansion described in Ref. [30], we consider $1/N_c$ as an expansion parameter and investigate the limiting behavior of the system when $N_c \to \infty$, neglecting higher-order with respect of $1/N_c$ terms. In this limit, the expected values of operators grow with $N_c$ faster than quantum corrections to these expected values. This enables us to transform Eqs. (2)–(6) to a system of equations for c-numbers in the leading order of $1/N_c$, while quantum corrections can be presented as a classical noise in the second order of $1/N_c$.

Since the cell size is smaller than the wavelength of radiation, we assume that the Rabi frequency is the same for all the atoms in the cell. For $i$-th mode and $k$-th cell $\Omega_{ij} = \Omega_{ik}$ for any $j$-th atom from the cell. The corresponding terms in Eqs. (2) and (3) become $\sum_{j \in cell_k} \Omega_{ij} \hat{\sigma}_j = \Omega_{ik} N_c \hat{J}_k$. In the end, we obtain the system of equations for $c$-numbers, in which the unknown variables are denoted by the same letters as the corresponding operators except for operators $\hat{a}_i^\dagger$ and $\hat{J}_k^\dagger$. The latter operators are replaced by the $c$-number $a_{cc}^i$ and $J_{cc}^k$. To distinguish operators from the corresponding $c$-numbers, the subscripts of operators are replaced by superscripts for $c$-numbers:

$$\frac{d}{dt} a^i = \left(-\gamma_a/2 - i\Delta_i\right) a^i - i \sum_k \Omega_{ik} N_c J^k, \tag{12}$$

$$\frac{d}{dt} a_{cc}^i = \left(-\gamma_a/2 + i\Delta_i\right) a_{cc}^i + i \sum_k \Omega_{ik} N_c J_{cc}^k, \tag{13}$$

$$\frac{d}{dt} J^k = -J^k (\gamma_p + \gamma_D + \gamma_{deph})/2 + i \sum_i \Omega_{ik} a^i J_D^k + F_J^k, \tag{14}$$

$$\frac{d}{dt} J_{cc}^k = -J_{cc}^k (\gamma_p + \gamma_D + \gamma_{deph})/2 - i \sum_i \Omega_{ik} a_{cc}^i J_D^k + F_{Jcc}^k, \tag{15}$$

$$\frac{d}{dt} J_D^k = -(\gamma_p + \gamma_D)\left(J_D^k - D_0\right) + 2i \sum_i \Omega_{ik} \left(a_{cc}^i J^k - a^i J_{cc}^k\right) + F_{JD}^k. \tag{16}$$

The last terms in Eqs. (14)-(16) describe classical noises; their correlation functions are similar to the correlation functions for the operators, see Eqs. (8)-(11) [29].

In the absence of noise terms in Eqs. (14)-(16), Eqs. (12) and (13) as well as Eqs. (14) and (15) are complex conjugated. If initially, $a_{cc}^i(t=0) = \left(a^i(t=0)\right)^*$ and $J_{cc}^k(t=0) = \left(J^k(t=0)\right)^*$, then these variables remain complex-conjugated all the time. Since operators $\hat{a}_i^\dagger$, $\hat{a}_i$ and $\hat{J}_k^\dagger$, $\hat{J}_k$ do not commute, in Eqs. (14)-(15), one needs to consider different realizations for noises. Consequently, noises, while retaining correct correlators, are no longer complex conjugated [29, 30]. Then, Eqs. (12) and (13) and Eqs. (14) and (15) are also not complex conjugated. As a result, the variables $a_{cc}^i$, $a^i$ and $J_{cc}^k$, $J^k$ corresponding to the operators $\hat{a}_i^\dagger$, $\hat{a}_i$ and $\hat{J}_k^\dagger$, $\hat{J}_k$ ceased to be complex conjugated.

We use the Eqs. (12)-(16) to find the radiation output, the spectrum and the second-order coherence function of ASE and laser radiations.

## 3. AMPLIFIED SPONTANEOUS EMISSION

As mentioned above, the main characteristic that allows one to experimentally distinguish coherent radiation from a laser and incoherent black-body radiation is the second-order coherence function $g^{(2)}(\tau)$. The value of $g^{(2)}(\tau)$ for any radiation source tends to unity as $\tau \geq 1/\Delta\omega$ where $\Delta\omega$ is the radiation linewidth of the source. Consequently, the response time $\tau_d$ of an experimental setup should be smaller than $1/\Delta\omega$. If $\tau_d > 1/\Delta\omega$ the measured value of $g^{(2)}(0)$ is always about unity. The ideal situation is to select a single mode and reach $\Delta\omega = 0$. Experimentally, a spectrum of the investigated source is narrowed by filtering. For this reason, we study the second-order coherence function at a fixed frequency, $g^{(2)}(\omega,\tau) = \langle I_\omega(t)I_\omega(t+\tau)\rangle / \langle I_\omega(t)\rangle^2$, where $I_\omega$ is output intensity at the frequency $\omega$. Using $\hat{\mathbf{E}}(\mathbf{r},t) = \sum_i \mathbf{E}_i(\mathbf{r})\hat{a}_i(t)$ and $I(\mathbf{r},t) = \langle \hat{\mathbf{E}}^\dagger(\mathbf{r},t)\hat{\mathbf{E}}(\mathbf{r},t)\rangle$ we obtain the expression for $g^{(2)}(\omega_i,\tau)$:

$$g^{(2)}(\omega_i,\tau) = \frac{\langle \hat{a}_i^\dagger(t_{st})\hat{a}_i^\dagger(t_{st}+\tau)\hat{a}_i(t_{st}+\tau)\hat{a}_i(t_{st})\rangle}{\langle \hat{a}_i^\dagger(t_{st}+\tau)\hat{a}_i(t_{st}+\tau)\rangle\langle \hat{a}_i^\dagger(t_{st})\hat{a}_i(t_{st})\rangle}. \qquad (17)$$

In our computer simulations, we study the quantity

$g^{(2)}(\omega_i,0) = \langle a_{cc}^i(t)a_{cc}^i(t)a^i(t)a^i(t)\rangle / \langle a_{cc}^i(t)a^i(t)\rangle^2$ by numerically solving Eqs. (11)-(15).

As a model for ASE source, we use 1D waveguide of the length $L_0 = 1800\lambda$ that contains the region with the size $L = 140\lambda$ filled with gain medium atoms [see Fig. 1]. This region is pumped to create a positive population inversion. For brevity, we use $\lambda$ and the reversed transition frequency $1/\omega_{TLS}$ as units of length and time, respectively. The gain medium is divided into subwavelength simulation cells of the size $\Delta x = 0.1\lambda$ (the total number of cells is 1400). The EM field is modeled by 400 modes with equidistant frequencies within the interval $(0.94, 1.06)$. The parameters are chosen to be close to a gain medium based on dye R101. Relaxation rates are $\gamma_a = 2\times 10^{-3}\,\omega_{TLS}$, $\gamma_{deph} = 5.2\times 10^{-2}\,\omega_{TLS}$, and $\gamma_D = 10^{-6}\,\omega_{TLS}$, and the pump rate varies within the interval $\gamma_p = (2\times 10^{-7} - 2\times 10^{-5})\,\omega_{TLS}$. The coupling constant is $\Omega_{ik} = 1.8\times 10^{-6}\,\omega_{TLS}\cos(2\pi x_k/\lambda_i)$, where $x_k$ is the coordinate of the $k$-th cell, and $\lambda_i$ is the wavelength of the $i$-th mode.

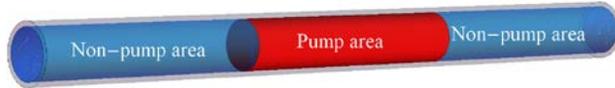

Fig. 1. The sketch of the ASE source based on a single-mode waveguide. The active medium in the central area of the waveguide (the red area) with the length $L$ is pumped by an external source; the active medium outside of this area is not pumped.

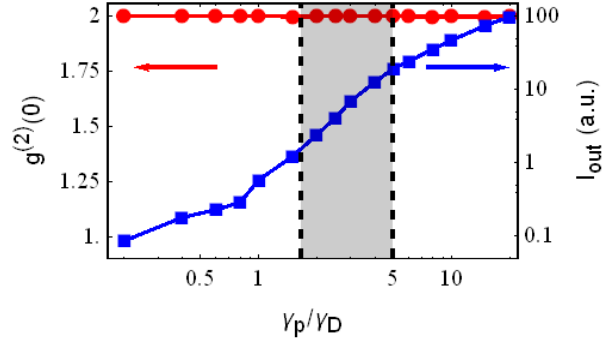

Fig. 2. The intensity of the EM field at the boundary of the active medium (the blue line) and the second-order coherence function, $g^{(2)}(0)$ (the red line). The left vertical dashed line shows the compensation threshold, at which pumping compensates for losses in the active medium. The right vertical dashed line represents the pump rate, at which the system passes into the nonlinear regime due to substantial saturation of the active medium. In this regime, the output power linearly depends on the pump power [25]. The shaded area shows the transitional region in which an increase in the pump rate results in an exponential increase of the radiation intensity.

Our numerical simulation confirms that the input-output curve of the ASE source has an s-shape (the blue line in Fig. 2) similar to the intensity of conventional laser radiation. The s-shape of the input-output curve of ASE sources arises due to the existence of three parameter regimes. At the low pump power, the propagation losses in the waveguide exceed the gain of the active medium. In this regime, the EM waves exponentially decay by propagation through the structure, and the output power linearly increases with the growth of the pump power. At the higher pump power, the gain exceeds the propagation losses and the EM waves exponentially intensify while the propagating, but the saturation of the active medium is inessential. Thus, the output power exponentially depends on the pump power (see shaded area in Fig. 2). The further increase of pump leads to saturation of the active medium, and the dependence of the output power on the pump power again becomes linear. The transitional region between first and third regimes is referred to as the ASE threshold [25].

Though within the transition range, the output intensity increases by more than an order of magnitude (see blue line in Fig. 2) the value of the second-order coherence function of ASE is independent of the pump rate (see the red line in Fig. 2) and is about 2, that is much greater than that for coherent light where $g^{(2)}(0)=1$.

In Figs. 3 (a), (c) and (b), (d), the spectral distribution of the photon number $n(\omega)$ and $g^{(2)}(\omega,0)$ are shown below and above the exponential part of the s-curve, respectively. For small pump rates, the spectrum of the system is similar to the spectrum of a single atom of the active medium [Fig. 3 (a)]. For pump rates that are higher than that of the transitional regime, the spectrum of the system markedly narrows [Fig. 3 (c)]. The observed narrowing is a common feature for both lasers and ASE.

At the same time, we can see that in spite of narrowing of the spectrum line, for any mode, $g^{(2)}(\omega,0) \approx 2.0$ independently of the frequency [Figs. 3 (b) and (d)]. As we discuss below, this distinguishes ASE from laser radiation.

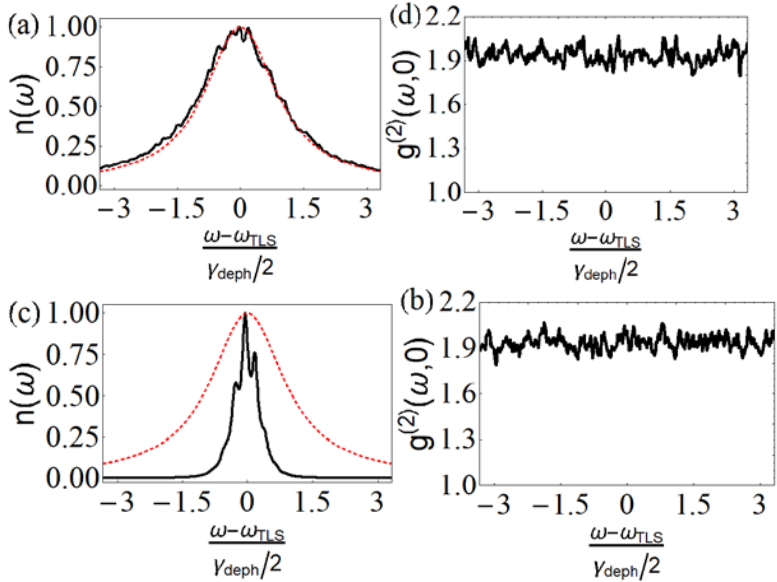

Fig. 3. Normalized spectra and $g^{(2)}(\omega,0)$ (black curves) of the ASE source for pump rates of $\gamma_D$ [Figs. (a) and (b)] and $20\gamma_D$ [Figs. (c) and (d) which correspond to the rates below and above the exponential part of the s-curve, respectively. Red curves show the absorption line of the medium. The spectrum is found by calculating the Fourier integral of the evolution function of the average $\langle a^i_{cc}(t_{st}+\tau) a^i(t_{st}) \rangle$. $g^{(2)}(\omega,0)$ is found by using Eq. (17). One frequency unit equals to the half width at the half height of the absorption line.

Thus, as Figs. 2 and 3 (b), (d) show, $g^{(2)}(0)$ does not depend on neither the pump rate nor the frequency, and it is about 2. This means that coherent properties of the ASE are distinguished from radiation of lasers; they are rather close to the black-body radiation.

## 4. LASER EMISSION

To test our results, we add mirrors to the system shown in Fig. 1 and make calculations similar to that discussed above. Thanks to mirrors, positive feedback would be formed, and the system would lase. In a laser, the value of $g^{(2)}(0)$ depends on the pump rate [31]: below the generation threshold, $g^{(2)}(0)$ must be close to two, above the threshold, it should tend to one. Our computer simulation confirms this assumption.

In Fig. 4, the dependence of the system output intensity on the pump rate is shown. This dependence has a pronounced s-shape (the blue line). The laser generation threshold (an inflection on the s-shaped curve) arises due to three regimes of a laser. For small pump rates, the amplification of radiation in the gain medium is not sufficient for the loss compensation in the waveguide and mirrors. In this regime, auto-oscillations are not established. For intermediate pump rates, in the system, auto-oscillations begin, but the nonlinear regime is not reached yet. The beginning of auto-oscillations is defined by the Maxwell-Bloch threshold in a system without noise. At a further increase of the pump rate, the population inversion of the active medium becomes significant causing the transition of the dependence of the radiation intensity from exponential to linear.

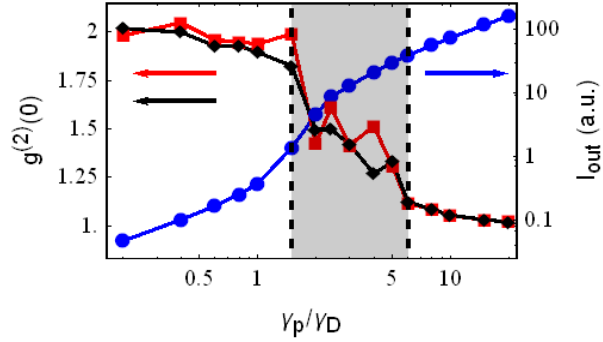

Fig. 4. The coherence function of two dominant modes (red and black curves) and the generation curve (the blue curve) of a laser. The curves are obtained as a result of solving the Maxwell-Bloch equations with noise. The left vertical dashed line shows the lasing threshold given by the Maxwell-Bloch equations without noise. The right vertical dashed line corresponds to the pump rate starting with which $g^{(2)}(0)-1$ is inverse-proportional to the average number of photons. Grey shading marks the transitional regime of the laser. The parameters of the active structure are the same as for the ASE system in Fig. 3. The amplitude reflectance of the mirrors is 0.8.

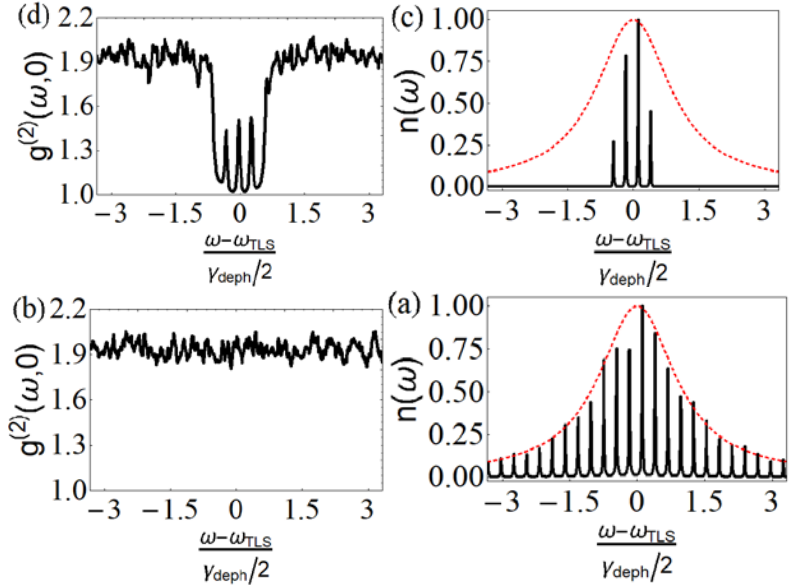

Fig. 5. Spectra and $g^{(2)}(\omega,0)$ of laser emission for pump rates $\gamma_p = \gamma_D$ (below the lasing threshold) and $\gamma_p = 20\gamma_D$ (above the laser threshold).

Below the generation threshold, the system spectrum exhibits a large number of lines corresponding to modes of the Fabry-Perot cavity [Fig. 5(a)]. The value of $g^{(2)}(0)$ for all modes is approximately 2 [Fig. 5(b)]. This is close to $g^{(2)}(0)$ for the system without mirrors [Fig. 3(b)]. Far above the generation threshold, there are four lines in the spectrum. These lines are at eigenfrequencies of the Fabry-Perot cavity that are the closest to the transition frequency of the active medium and consequently have the lowest values of thresholds. For these modes, $g^{(2)}(0)$ is close to 1 even though for all other modes it remains about 2.

Near the generation threshold, our model, Eqs. (11)-(15), give large fluctuations and the large spread in $g^{(2)}(0)$. This happens because, near the threshold, the laser behavior is similar to a second order phase transition [29, 30] in which a system undergoes strong fluctuations near the critical point [32]. In our case, the critical region corresponds to pump rates between $\gamma_p = 1.5\gamma_D$ and $6\gamma_D$ (the shaded area in Fig. 4).

Thus, in the system with mirrors, a pronounced generation threshold is observed. Our calculations show that below the threshold, for all modes, $g^{(2)}(0)$ significantly exceeds 1, while above the threshold, for the modes in which the generation occurs, $g^{(2)}(0)$ tends to unity. $g^{(2)}(0)$ decreases only for the eigenmodes of the system with mirrors and only when there is coherent feedback. This agrees with the expected behavior of a laser. We would like to emphasize that in this subsection we use exactly the same model as in the previous one except that mirrors that create a cavity are added. Then, using the same procedure as in the previous subsection, we obtain well-known results. This can serve as a confirmation of the correctness of the results obtained for ASE.

## 5. DISCUSSION AND CONCLUSIONS

In this paper, we consider statistical properties of ASE in a system with a pumped active medium without mirrors. We show that similar to a laser, in the input-output curve of ASE, there are three. However, in these regions the second-order coherence functions $g^{(2)}(0)$ exhibits different behavior for ASE and laser. In the first region, below the compensation and lasing thresholds, $g^{(2)}(0)$ of both systems is close to 2 In the second (transition) region (below the saturation threshold), $g^{(2)}(0)$ of ASE remains constant, while for a laser, $g^{(2)}(0)$ exhibits fluctuations. Finally, in the saturation region, $g^{(2)}(0)$ of ASE remains 2, while for lasers, $g^{(2)}(0)$ tends to unity.

Thus, above the threshold, the second-order coherence functions of ASE and a laser are different. This makes it possible to detect the positive feedback in a system. For example, it is known that in interstellar gases [33-35] and planetary atmospheres [36, 37] the amplification of EM waves by stimulated emission occurs. Such radiation sources are called astrophysical masers or lasers. In Refs. [35, 38], it is claimed that such sources operate as random lasers ($g^{(2)}(0) \approx 1$). An alternative mechanism could be the phenomenon of ASE [39] ($g^{(2)}(0) \approx 2$). An investigation of the second-order coherence function of radiation from an astrophysical "maser" could shed light on the mechanism of formation of such sources. In turn, knowledge of the mechanism of the formation of such anomalies can enable one to estimate the concentration of the scatters (molecules, atoms, or ions) in interstellar clouds.

In addition to its fundamental significance, the difference between ASE and laser radiations is of practical importance. For ghost imaging applications, light sources with super-Poissonian distributions of photons are necessary. Currently, either lasers with rotating grounded glasses [14-16], producing radiation with $g^{(2)}(0)$ between 1.25 and 1.9, or incoherent lamps with a frequency filter [17, 18] with $g^{(2)}(0)$ of about 1.05 are used as light sources. In this paper, we demonstrate that the second-order coherence function of ASE sources is about 2 that is closer to the coherence of lasers with rotating grounded glass. However, ASE sources are much simpler and easier to manufacture than lasers. This makes these promising light sources for ghost imaging [10].

**Funding**. National Science Foundation (NSF) (DMR-1312707).

**Acknowledgment**. We thank V. I. Balykin for helpful discussions.